\documentclass{pre}
\newcommand{\mm}{MM}
\newcommand{\SC}{SC}
\newcommand{\etal}{\emph{et al.}}

\newcommand{\eg}{e.g.}
\newcommand{\ls}{$\ell \cdot s$}
\newcommand{\SHF}{SHF}
\newcommand{\asurf}{\mbox{$a_{\rm surf}$}}
\newcommand{\avol}{\mbox{$a_{\rm vol}$}}
\newcommand{\asym}{\mbox{$a_{\rm sym}$}}
\renewcommand{\vec}[1]{{\bf #1}}
\begin{document}
\title{Extrapolation of mean--field models\\ 
       to superheavy nuclei} 
\author{Michael Bender}
\address{Gesellschaft f\"ur Schwerionenforschung,
         Darmstadt, Germany}
\maketitle
%
%
\abstracts{
The extrapolation of self--consistent nuclear mean-field
models to the region of superheavy elements is discussed with 
emphasis on the extrapolating power of the models. The predictions
of modern mean-field models are confronted with recent experimental 
data. It is shown that a final conclusion about the location of the 
expected island of spherical doubly-magic superheavy nuclei cannot
be drawn on the basis of the available data.
}
%
%
\section{Introduction}
\label{sect:intro}
\footnotetext{Invited talk at the International Conference on 
``Fusion dynamics at the extremes'', Dubna, Russia, May 25--27, 2000.
}
The last decade brought with the synthesis of superheavy nuclei with 
\mbox{$Z=110$}--$112$ at GSI (Darmstadt) and JINR (Dubna)
a renewal of the interest in the properties of superheavy nuclei%
\cite{SHReview1,SHReview2}. These are by definition those nuclei 
with \mbox{$Z > 100$} which have a negligible liquid--drop fission
barrier and are stabilized by quantal shell effects only. 
The ultimate goal is to reach an expected ``island of stability'' 
located around the next spherical doubly--magic nucleus which was 
predicted to be $^{298}_{184}114$ thirty years 
ago\cite{Meldner,Meldner2,Nilsson}.
Recent experiments performed at JINR Dubna\cite{Dubna} give evidence 
for the synthesis of the neutron--rich nuclides 
\mbox{$^{283}112$}, \mbox{$^{287-289}114$}, \mbox{$^{292}116$}, 
while at Berkeley three $\alpha$--decay chains attributed to the even
heavier ${}^{293}118$ were observed\cite{118175}. The measured 
$\alpha$--decay chains of these new nuclides turn out to be consistent 
with theoretical predictions\cite{Smo99a,Cwi99a,Ben00a}.
While earlier superheavy nuclei could be unambiguously identified by 
their $\alpha$--decay chains leading to already known nuclei, the decay
chains of the new--found superheavy nuclei cannot be linked to any 
known nuclides. The new discoveries still have to be viewed carefully, 
see \eg\ the critical discussion in\cite{Arm00a}. 

The recent experimental developments are accompanied by a significant 
progress in the theoretical modeling of superheavy nuclei by means of 
nuclear mean--field (MF) models. MF models can be divided into two 
major classes, (i) self--consistent (SC) ones where the single-particle 
wave functions are calculated from an average nuclear potential which 
in turn depends on these wave functions and (ii) 
macroscopic--microscopic (MM) models which are composed by a generalized 
liquid--drop model that governs the bulk properties and a 
single--particle potential from which the shell correction is derived.

Recent theoretical work reveals systematic differences among the
predictions of the models. While modern refined \mm\ models confirm 
the older prediction of $^{298}_{184}114$ for the next spherical 
doubly--magic nucleus, nearly all \SC\ models 
shift that property to higher charge numbers, depending on the class
of \SC\ models to either $^{292}_{172}120$ or $^{310}_{184}126$.
The reasons for these conflicting predictions and their implications 
are the topic of this contribution.
%
%
\section{Models}
\label{sect:models}
Most \SC\ mean--field models can be viewed as energy density theories in 
the spirit of the Hohenberg--Kohn--Sham (HKS) approach\cite{Hoh64a,Koh65a}  
originally introduced for many--electron systems which
is nowadays a standard tool successfully applied in atomic, 
molecular, cluster, and solid--state physics. 
Starting point is the existence theorem\cite{DFT} for a unique energy
functional ${\cal E}$ depending on all local densities and currents 
that can be constructed from the general single--particle density matrix 
\begin{equation}
\label{eq:densitymatrix}
\hat\rho
\equiv \rho (\vec{r}, \sigma, t; \vec{r}', \sigma', t')
= \sum_k v_k^2 \; \psi_k^* (\vec{r}', \sigma', t') \;
               \psi_k (\vec{r}, \sigma, t)
\end{equation}
which gives the exact ground--state energy of a system of identical 
Fermions when ${\cal E}$ is calculated for the exact ground--state 
density $\hat{\rho}$. $\vec{r}$, $\sigma$, and $t$ are the spatial, 
spin, and isospin coordinates of the wave functions $\psi_k$. 
The HKS approach maps the nuclear many--body problem for the ``real''
highly--correlated many--body wave function onto a system of 
independent particles in effective, so--called Kohn--Sham orbitals
$\psi_k$. The equations of motion of the $\psi_k$ are derived from 
a variational principle
\begin{equation}
\delta {\cal E}
= 0 
\quad \Rightarrow \quad
\hat{h} (\vec{r}, \sigma, t) \, \psi_k (\vec{r}, \sigma, t)
= \epsilon_k \, \psi_k (\vec{r}, \sigma, t)
\end{equation}
where the single--particle Hamiltonian $\hat{h}$ is the sum of the
kinetic term $\hat{t}$ and the self--consistent potential $\Gamma$
that is calculated from the actual density matrix
\begin{equation}
\hat{h}
= \frac{\delta {\cal E}}{\delta \hat\rho}
= \hat{t} + \hat\Gamma [\hat\rho]
\quad .
\end{equation}
The existence theorem for the energy functional, however, makes no 
statement about the actual structure of the effective interaction.
Guided by symmetry principles and phenomenological knowledge about 
nuclei, the aim is to find the most simple energy functional 
which incorporates all relevant physics 
and to adjust its parameters uniquely to a selected set of nuclear key data.
The two most widely used \SC\ models are the (non--relativistic)
Skyrme--Hartree--Fock (\SHF) model and the relativistic mean--field 
(RMF) model.
The \SHF\ energy functional contains all bilinear combinations of 
local densities that are invariant under rotational, translational, 
parity, and time--reversal transformations up to second order in the 
derivatives plus a simple density dependence\cite{Dob96a}. 
The standard RMF energy functional assumes the nucleus to be a system 
of Dirac nucleons interacting via scalar and vector fields usually 
associated with $\sigma$, $\omega$, and $\rho$ mesons, again plus a 
simple density dependence\cite{Rei89a}.
Pairing is treated in both SHF and RMF using the same (non--relativistic)
local pairing energy functional corresponding to a delta pairing 
force\cite{Ben99a}.

\mm\ models can be motivated as an approximation to \SC\ models by
means of the Strutinsky theorem\cite{Str67a,Nilssonbook}. 
The binding energy is separated into a large average part 
$\tilde{E}$ depending smoothly on $N$ and $Z$ and a small shell 
correction $E_{\rm shell}$ that describes local fluctuations of the 
binding energy caused by variations of the density of single--particle 
levels around the Fermi surface
\begin{equation}
\label{eq:mm}
E (Z,N)
= \tilde{E} (Z,N) + E_{\rm shell} (Z,N)
\end{equation}
For superheavy nuclei $\tilde{E}$ is of the order $-2000$ MeV while
$E_{\rm shell}$ fluctuates in the range $-15 \; {\rm MeV} \; \leq 
E_{\rm shell} \leq +15 \; {\rm MeV}$.
In \mm\ models the self--consistent coupling of shell 
structure and bulk properties is replaced by (independent but similar) 
parameterizations of the the density distribution and the
single--particle potentials with $N$ and $Z$.
Modern \mm\ models combine a finite--range liquid--drop (YPE) or droplet 
(FRDM) model for $\tilde{E}$ and a phenomenological single--particle model 
based on either the Woods--Saxon (WS) or the Folded--Yukawa (FY) potential 
for the calculation of $E_{\rm shell}$.
The two most widely used models are the FRDM+FY\cite{Mol94a}
and the YPE+WS models\cite{Smo97a}. 
It is to be noted that
loosely--bound systems like superheavy nuclei where the Fermi energy 
is close to the continuum require a more careful treatment of unbound 
states than done in the standard approach used in large--scale calculations 
with \mm\ models\cite{Kru00a}. 
%
%
\section{Nuclear exotica in superheavy nuclei}
In superheavy nuclei the repulsive Coulomb interaction is not
counteracted by the surface tension which leads to a 
vanishing liquid--drop fission barrier. Therefore
the Coulomb field cannot be treated as a small perturbation 
atop the nuclear mean field, it pushes the protons 
to the nuclear surface which is counteracted by the symmetry 
energy restoring a similar distribution of protons and neutrons 
and the density dependence of the effective interaction restoring 
the saturation density of nuclear matter (for given 
asymmetry). With that superheavy nuclei probe the balance of 
bulk properties of effective interactions.
At the same time the large density of single--particle states provides 
a sensitive probe for even subtle details of the single--particle 
structure. Again the Coulomb potential induces 
significant changes in the proton shell structure, compared to lighter
nuclei single--particle states with large angular momentum $j$ 
are lowered compared to small--$j$ states\cite{Cwi96a}.

Up to now most of the understanding of phenomena in superheavy nuclei 
was obtained on the basis of \mm\ models.
Recent work employing \SC\ models, however, predicts some new 
phenomena which cannot be consistently described by (current) 
\mm\ models. Examples are exotic radial density distributions 
like ``semi--bubbles''\cite{Dec99a} (which are essential
for the appearance of spherical shell closures at \mbox{$Z=120$} and 
\mbox{$N=172$} predicted by some \SC\ models\cite{Ben99a})
and the variation of the surface diffuseness in
superheavy nuclei. It is well--known that the appearance of a proton 
shell at \mbox{$Z=126$} requires a larger surface diffuseness 
than assumed in the standard parameterizations of the FY and WS 
single--particle potentials\cite{Mye98a,Mye99a}.
Standard \mm\ models use a parameterization of the radial shape of 
the density distribution and single--particle potentials that has no
free parameters to minimize the binding energy and therefore are not
flexible enough to incorporate both of these effects. It can be expected
that these models loose their validity for nuclei with large charge number.
\SC\ models have to be preferred when describing these phenomena
as they make no assumption at all on the profile of the 
density distributions of protons and neutrons.
Besides the limitations of the currently used parameterizations 
the deeper reason why these phenomena cannot 
be easily described by \mm\ models is that they are caused by the
coupling of single--particle degrees of freedom and bulk properties
which becomes more pronounced for loosely bound systems.
%
%
\section{Fits and Parameterizations}
\label{sect:fits}
Although they might not contain all physics relevant for superheavy
nuclei beyond the known region current \mm\ models give a better 
description of binding energies than the best \SC\ models\cite{Mye99a}. 
This is not too surprising as the aims and the 
fit strategies of \mm\ and \SC\ models are very different: \mm\ models
are optimized for the description of masses by adjusting the parameters 
of the macroscopic part $\tilde{E}$ to all known masses.
The density distribution entering the macroscopic part of the model is 
parameterized as a function of $N$ and $Z$ to reproduce the global
systematics of radii, the microscopic potential is adjusted to
reproduce systematics of single--particle spectra throughout the chart
of nuclei. On the other hand, \SC\ models are designed to give
a consistent description of nuclear ground states, single--particle
spectra, collective excitations like giant resonances and rotational 
bands, and large--amplitude collective motion within the same model. 
The parameters of the interaction are usually adjusted to masses and 
data on the charge distribution of selected spherical nuclei, and 
in some cases to selected data on infinite nuclear matter (INM) to 
compensate for the small number of sample points in the fit.
As the description of single--particle spectra and bulk properties 
cannot be separated the interactions need not to be adjusted to 
spectral data with the exception of 
the spin--orbit (\ls) interaction. As a purely relativistic 
effect it is naturally incorporated in the RMF\cite{Rei89a} 
which reproduces data on \ls\ splittings without being 
adjusted to any data on single--particle spectra at all\cite{Ben99a}.
In non--relativistic models -- either \mm\ or \SC\ ones --
the \ls\ force has to be put in by hand and needs to be adjusted 
to spectral data.

A detail to be kept in mind when comparing masses from 
\mm\ and \SC\ models is that in the fit of \mm\ models the 
absolute error of ${\cal E}$ is minimized, while for \SC\ models 
usually the relative error of ${\cal E}$ is minimized\cite{Fri86a}
which allows for much larger absolute errors in heavy nuclei, up to 
5.5 MeV in $^{208}$Pb corresponding to $0.35\%$.

Besides the technical reason that everything else than a fit to a
small sample of spherical nuclei is too time--consuming there are 
also physics reasons for the usual small sample of fit nuclei. The 
many--body wave function $\Phi$ of mean--field models breaks symmetries 
which are obeyed by the effective energy functional and the ``exact'' 
wave function. An instructive example for this ``symmetry dilemma'' 
is violation of translational symmetry which is 
unavoidable as the center--of--mass (c.m.) of the nucleus is localized 
by the mean--field potential. Although $\Phi$ has vanishing total 
momentum $\langle \vec{P} \rangle = 0$ it is not an eigenstate of 
the momentum operator, $\langle \vec{P}^2 \rangle \neq 0$. 
This means that $\Phi$ is not the pure ground state but contains an 
admixture of excited states with finite momentum which decreases the  
calculated mass. Rigorous restoration of the broken symmetry by 
means of projection is too costly to be used in large--scale 
calculations. The method of choice is to estimate the contribution
from excited states to the calculated energy and subtract this c.m.\ 
correction $E_{\rm c.m.}$ to obtain the binding energy. An unexpected 
side--effect is that the nuclear matter properties of effective 
interactions depend on technical details of the actual c.m.\ correction 
performed during the fit as the effective interaction has to compensate 
for the difference between approximations and the exact value of 
$E_{\rm c.m.}$. 
Often--used simple schemes for c.m.\ correction lead accidentally 
to too large a surface coefficient\cite{Ben00b} which has visible impact 
on the extrapolation of the models to heavy systems and large deformation.
A similar influence on the results of a fit can be expected for 
other corrections for spurious motions, namely the rotational and 
vibrational correction. Luckily their contribution to the binding 
energy can be suppressed by choosing spherical nuclei with stiff 
potential energy surfaces\cite{Fri86a} which explains the usual
small sample of fit nuclei.

There are numerous parameterizations of the SHF and RMF models to be 
found in the literature. 
Results discussed here are obtained with the SHF 
interactions SkM*\cite{SkM*}, SkP\cite{SkP}, SLy6\cite{SLyx}, 
SkI3 and SkI4\cite{Rei95a}, and the RMF forces NL3\cite{NL3}, 
NL--Z\cite{NLZ}, and NL--Z2\cite{Ben99a}. SkP uses effective mass 
\mbox{$m^*/m = 1$} and is designed to describe both mean--field and pairing 
interaction. The other Skyrme forces all have smaller effective masses 
around \mbox{$m^*/m \approx 0.75$}. SkM$^{*}$ was 
first to deliver acceptable incompressibility and fission
properties. SLy6 stems from an attempt to cover properties 
of pure neutron matter together with normal nuclear ground--state
properties, while SkI3 and SkI4 stem from a recent 
fit including data from exotic nuclei and use 
a variant of the Skyrme parameterization where the \ls\ force 
is complemented by an explicit isovector degree--of--freedom\cite{Rei95a}. 
They are designed to overcome the different isovector trends
of \ls\ coupling between conventional Skyrme forces and the RMF.
The RMF parameterizations NL--Z, NL--Z2, and NL3 use the standard
non--linear \emph{ansatz} for the RMF model. NL--Z aims at a best fit 
to nuclear ground--state properties. NL--Z2 matches exactly the same 
enlarged set of data like the SkI$x$ forces. NL3 results 
from a fit including neutron rms radii and nuclear matter data. 

%
%
\section{Predictive Power}
\label{sect:predict}
%
%
\begin{table}[t]
\parbox[t]{6cm}{
\begin{center}
\begin{tabular}{lccc}
\hline\noalign{\smallskip}
Force & \protect\avol
      & \protect\asym   
      & \protect\asurf \\
      & $[{\rm MeV}]$
      & $[{\rm MeV}]$ 
      & $[{\rm MeV}]$  \\
\noalign{\smallskip}\hline\noalign{\smallskip}
SkM*   & $-15.8$ & 30.0 & 17.6 \\
SkP    & $-15.9$ & 30.0 & 18.0 \\ 
SkI3   & $-16.0$ & 34.8 & 17.5 \\ 
SkI4   & $-15.9$ & 29.5 & 17.3 \\
SLy6   & $-15.9$ & 32.0 & 17.4 \\ 
\noalign{\smallskip}\hline\noalign{\smallskip}
\end{tabular}
\end{center}
}
\parbox[t]{6cm}{
\begin{center}
\begin{tabular}{lccc}
\hline\noalign{\smallskip}
Force & \protect\avol
      & \protect\asym   
      & \protect\asurf \\
      & $[{\rm MeV}]$
      & $[{\rm MeV}]$ 
      & $[{\rm MeV}]$  \\
\noalign{\smallskip}\hline\noalign{\smallskip}
NL--Z  & $-16.2$ & 41.7 & 17.7 \\
NL--Z2 & $-16.1$ & 39.0 &      \\  
NL3    & $-16.2$ & 37.4 & 18.5 \\
\noalign{\smallskip}\hline\noalign{\smallskip}
\end{tabular}
\end{center}
\vfill
}
\caption{\label{Tab:NucMat}
Compilation of nuclear matter properties for a number of typical 
parameter sets. SkM*--SLy6 are Skyrme forces, and NL--Z, NL--Z2,
and NL3 RMF forces. \protect\avol\ denotes the volume coefficient or 
energy per nucleon, \protect\asym\ the (volume) asymmetry coefficient, 
and \protect\asurf\ the surface coefficient. Empirical values for the
volume coefficients derived from the liquid--drop model are 
\mbox{$\avol = -16.0 \pm 0.2$} and \mbox{$\asym = 32.5 \pm 0.5$}.
See\protect\cite{Ben00b} for more details.
}
\end{table}
%
%
Table~\ref{Tab:NucMat} summarizes those nuclear matter properties that
can be directly linked to the leading terms of the liquid--drop model
\begin{equation}
\label{eq:LDM}
E_{\rm LDM}
=  \avol \, A
  + \asym \, I^2 \, A  
  + \asurf \, A^{2/3}
  + \ldots
\end{equation}
for some typical SHF and RMF forces. Most non--relativistic 
(relativistic) interactions agree among each other in the values 
for \avol\ and \asym,
but relativistic and non--relativistic interactions differ 
significantly. While the Skyrme force values are close to the 
empirical ones, the RMF forces give larger \avol\ and \asym.
work much better in that respect. It is important to note,
Not all values for Skyrme interactions are predictions, for the 
SLy6, SkP, and SkM* forces nuclear matter properties 
were used as input data during their fit. As already mentioned above 
the predicted value for \asurf\ is accidentally correlated to the scheme 
for c.m.\ correction used in the fit of the interaction\cite{Ben00b}, 
forces with simple c.m.\ correction like SkP and NL3 
have systematically larger \asurf\ than the others.
%
%
\begin{figure}[t!]
\centerline{\epsfig{file=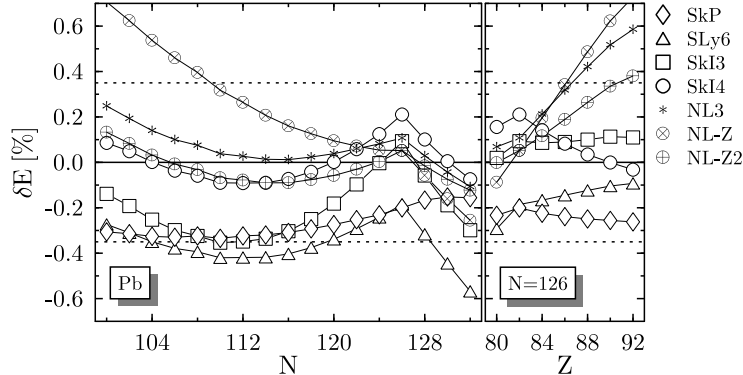}}
\caption{\label{Fig:deltaE}
Relative error $\delta E = (E_{\rm calc} - E_{\rm expt}) / E_{\rm expt}$
(in $\%$) of the binding energy for the chain of Pb isotopes and the
chain of $N=126$ isotones. Negative values denote
under--bound, positive values over--bound nuclei. The horizontal 
lines at $\pm 0.35 \%$ indicate the relative error in binding energy 
allowed for good fits\protect\cite{Rei89a,Rei95a}.
}
\end{figure}
%
%

\begin{figure}[t!]
\centerline{\epsfig{file=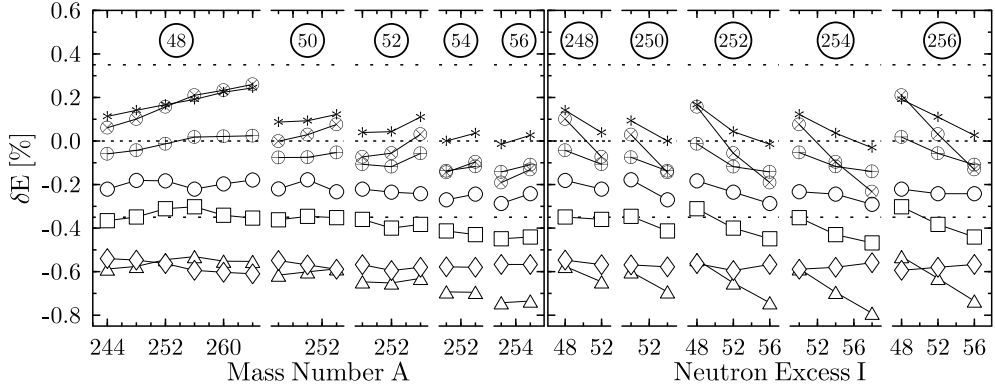}}
\caption{\label{Fig:deltaE2}
Relative error the binding energy for the heaviest deformed nuclei where 
the mass is known plotted for constant $I = N - Z$ against $A = N + Z$
(left panel) and vice versa (right panel) to separate trends 
in the isoscalar and isovector channels of the effective interactions.
Symbols as in Fig.~\protect\ref{Fig:deltaE}.
Data are taken from\protect\cite{Ben99b,Bue98a}.
}
\end{figure}

A first impression of the quality of mean--field models for finite
nuclei is provided
by the systematics of the relative error of 
binding energies, see Figs.~\ref{Fig:deltaE} and~\ref{Fig:deltaE2}
for typical results. For the chain of spherical Pb isotopes the results 
remain essentially within the bounds of $0.35 \%$ allowed in the fits, 
while the spherical $N=126$ isotones reveal already some unresolved 
trends in the RMF interactions. 
The slope of $\delta E$ corresponds to an error of the 
two--nucleon separation energies. Unlike modern \SHF\ interactions 
which have four or even five independent isovector terms, the RMF has 
one isovector term only which seems to be insufficient. The small change
in the isovector coupling between NL--Z and NL--Z2 improves visibly
the trends with $N$ but at the same time worsens the trends 
with $Z$. But one has to be careful as both $I$ and $A$ 
change along isotopic and isotonic chains which mixes trends
in the isovector and isoscalar channel. These can be separated 
plotting $\delta E$ against $I$ and $A$, see Fig.~\ref{Fig:deltaE2}
with data for superheavy nuclei.
It can be expected that in this extrapolation of the models the $\delta E$ 
spread more than in the case of the lighter nuclei. It is
gratifying, however, to see that most of the interactions stay within 
or at least close to the bounds of $0.35 \%$ error.
When plotted versus $A$ one obtains essentially flat curves for all 
\SHF\ forces, while there is a small but visible slope for the RMF forces 
that points at an unresolved isoscalar trend. When plotted versus $I$
all forces (perhaps with the exception of SkP) show slopes 
which point at unresolved trends in the isovector channel, 
even for modern forces like NL3 or SLy6
fitted with bias on good isovector properties. One has to be
careful with the interpretation of Fig.~\ref{Fig:deltaE2} since the
range of known masses in $I$ is rather small and wrong trends 
in the ``macroscopic part'' of the \SC\ models might interfere with
local flaws in the microscopic part. An example for that is
the peak in the $\delta E$ around $^{208}$Pb in the left panel 
of Fig.~\ref{Fig:deltaE} which hints at an unresolved shell effect. 
It is to be noted that missing shell effects might also
correspond to correlations beyond the mean--field level, see
the example of c.m.\ correlations discussed in\cite{Ben00b}.
%
%
%
%

\begin{figure}[t!]
\centerline{\epsfig{file=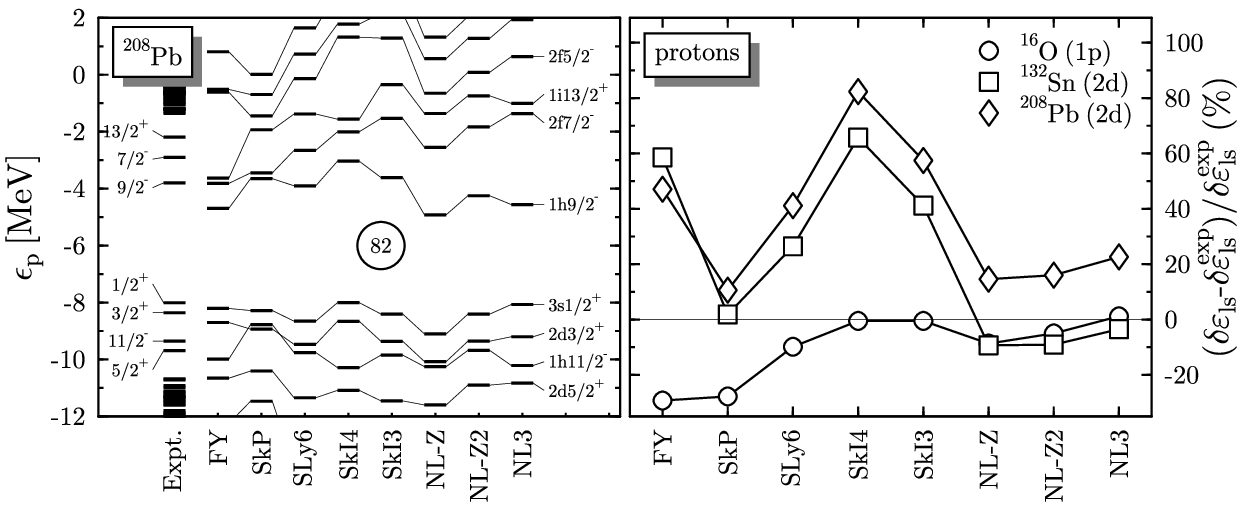}}
\caption{\label{Fig:spectra}
Single--particle spectra of the protons in $^{208}$Pb (left panel) 
and relative error of the spin--orbit splittings
\mbox{$\delta \epsilon_{\rm ls} 
= (\epsilon_{\rm ls,calc} - \epsilon_{\rm ls,expt}) / \epsilon_{\rm ls,expt}$}
(in $\%$) of proton states close to the Fermi surface in 
$^{16}$O, $^{132}$Sn and $^{208}$Pb for the interactions as indicated.
Negative errors denote calculated values which are too small.
FY denotes the folded--Yukawa single--particle potential widely used
in \protect\mm\ models. Data are taken from\protect\cite{Ben99a}.
}
\end{figure}

As there are no experimental data on spectral properties of 
superheavy nuclei, the predictive power of the models for 
single--particle spectra has to be examined looking at lighter 
nuclei, see Fig.~\ref{Fig:spectra}.
None of the existing parameterizations of \SC\ or \mm\ models is
able to give a proper description of the level ordering and the
spin--orbit splitting in heavy nuclei such as $^{132}$Sn or
$^{208}$Pb. Levels with large angular momentum are usually pushed
up too far in the single--particle spectrum and all non--relativistic
models (either \SC\ and \mm\ ones) show a wrong trend of the spin--orbit
splitting with $A$ and therefore usually overestimate the 
proton spin--orbit splitting in heavy nuclei. This is devastating 
for interactions where the spin--orbit coupling constant is adjusted 
to data for $^{16}$O. Including data on heavy nuclei in the fit gives
better overall agreement but cannot remove the overall wrong trend
as can be seen from the example of SkP. 
It is surprising that Skyrme forces with an additional degree of 
freedom in the spin--orbit interaction (SkI3, SkI4) perform worse 
in that respect than standard forces (SkP, SLy6). The overall 
performance of the RMF interactions is much better.
%
%
%
%
%
%
\begin{figure}[b!]
\centerline{\epsfig{file=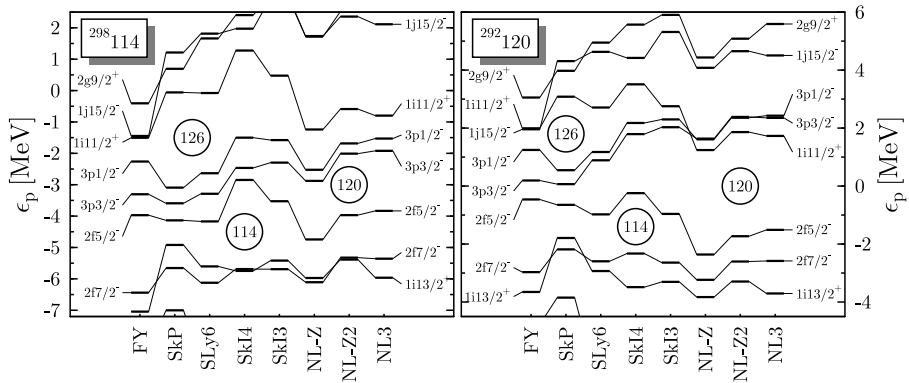}}
\caption{\label{Fig:spectra2}
Single--particle spectra of the protons in ${}^{298}_{184}114$ and
${}^{292}_{172}120$ at spherical shape for the interactions as indicated.
Note that in spite of the double shell closure the ground state 
of ${}^{292}_{172}120$ is actually deformed for most Skyrme 
interactions\protect\cite{Ben98a,Cwi00a}, while the strong \mbox{$N=184$}
shell stabilizes for most interactions the spherical shape of 
${}^{298}_{184}114$. Data are taken from\protect\cite{Ben99a}.
}
\end{figure}
%
%

As most mean--field models predict the same level--ordering in the 
superheavy region, already slight changes in the relative distances 
among the interactions lead to different magic numbers, see 
Fig.~\ref{Fig:spectra2} for typical results.
Large spin--orbit splitting as in case of the FY and SkI4 models favors 
\mbox{$Z=114$}, but as these interactions overestimate the spin--orbit 
splitting of proton states in heavy nuclei this prediction is very 
doubtful. Non--relativistic models in general prefer \mbox{$N=184$}
for neutrons, while small spin--orbit splitting in connection with 
``semi--bubble'' shapes leads to \mbox{$Z=120$} and \mbox{$N=172$} 
as it happens for SkI3 and the relativistic forces NL3, NL--Z, and NL--Z2. 
\SC\ interactions with large effective mass close 
to \mbox{$m^*/m \approx 1.0$} like SkP give an average level density 
so large that most of the shell effects are washed out and the proton 
shell is shifted to \mbox{$Z=126$}.
A common feature of spherical shells as predicted by \SC\ models is
their strong nucleon--number dependence\cite{Ben99a,Rut97a}:
While the spectra for the non--\SC\ FY model remain basically 
unchanged, the spectra from \SC\ models change dramatically going from 
${}^{298}_{184}114$ to ${}^{292}_{172}120$, the spherical shells at 
\mbox{$Z=120$}, \mbox{$Z=126$} and \mbox{$N=172$} are all restricted to 
a narrow range of $N$ and $Z$, a magic \mbox{$N=184$} often excludes 
a shell closure at \mbox{$Z=120$}.
%
%
\begin{figure}[t!]
\centerline{\epsfig{file=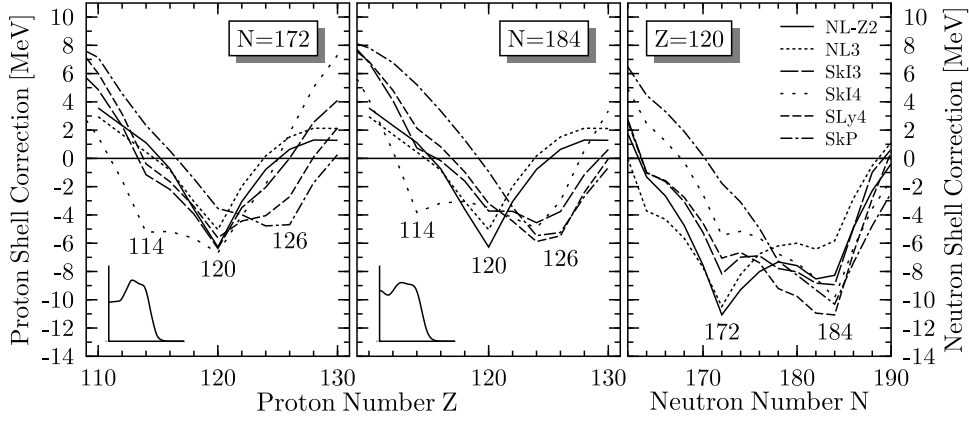}}
\caption{\label{Fig:sc:mic}
Proton shell correction extracted from calculated self--consistent 
binding energies of the \mbox{$N=172$} and \mbox{$N=184$} isotones (left
and middle panel) and neutron shell correction for the \mbox{$Z=120$}
isotopes. The same energy scale is used for all panels. Note that for 
the majority of Skyrme interactions all \mbox{$N=172$} isotones are 
predicted to be deformed. The inserts show the density profile for
$^{292}$120 (left panel) and $^{304}$120 (middle panel) calculated with 
NL--Z2. They demonstrate the strong neutron--number dependence of the 
``semi--bubble'' shapes which are responsible for the magic numbers 
\mbox{$Z=120$} and \mbox{$N=172$}. 
Data are taken from\protect\cite{Kru00a}.
}
\end{figure}
%
%

The appearance of gaps in the single--particle spectra alone is not 
sufficient to stabilize a superheavy nucleus. Much more important 
is the extra binding and its strong shape dependence obtained from 
a smaller than average level density. The shell correction 
can be viewed as the natural measure for this ``shell effect'' and
provides a powerful tool to analyze also results obtained in fully 
self--consistent calculations, see Fig.~\ref{Fig:sc:mic}.
It has to be emphasized that the shell correction is not equivalent 
to the gap in the single--particle spectrum and that a quantitative 
comparison of the two quantities cannot be made. 
For \mbox{$N=172$} most of the Skyrme forces give a magic \mbox{$Z=120$}, 
while for \mbox{$N=184$} the minimum is shifted to \mbox{$Z=124$}--126
in all cases. RMF forces give consistently different results, the 
strongest shell effect is at \mbox{$Z=120$} independent on the neutron 
number. Most interactions predict also a broad region of negative neutron 
shell  correction. Again a systematic difference between 
the models: while the RMF points at \mbox{$N=172$} as the dominant neutron 
shell, all SHF forces predict \mbox{$N=184$}. Note that shell corrections 
for neutrons are much larger than for protons. This partly explains the 
finding that spherical ground states of superheavy nuclei are usually 
correlated with magic neutron numbers, not proton 
numbers\cite{Cwi96a,Bue98a,Lal96a}.
All this suggests that the ``island of stability'' is not so much coupled 
to particular shell closures but a region of nuclei with low level density
which is also found in \mm\ models\cite{Mol94a}.
Fig.~\ref{Fig:sc:mic} demonstrates also that it is nearly impossible to 
discriminate between models giving different shell closures by looking at 
systematics of binding energies or $Q_\alpha$ values alone. Although the 
proton shell correction might be peaked at different points, its variation 
is often too small to be visible within the uncertainty of \SC\ models.
%
%
\begin{figure}[t!]
\centerline{\epsfig{file=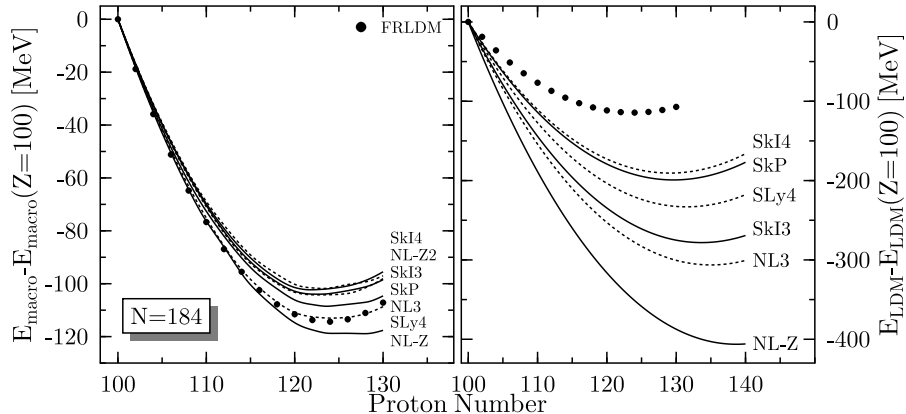}}
\caption{\label{Fig:sc:mac}
Macroscopic energy $\tilde{E}$ extracted from calculated self--consistent 
binding energies (left panel) and macroscopic energy $E_{\rm LDM}$ 
from the LDM expansion (\protect\ref{eq:LDM}) of the binding energy
(right panel) for the chain of \mbox{$N=184$} isotones. 
The phenomenological macroscopic energy from the YPE model is given for
comparison. Note that the scales of the two panels are very different.
To illustrate the $Z$ dependence all energies are normalized
to the YPE value at \mbox{$Z=100$}. 
Data taken from Ref.\protect\cite{Kru00a}.
}
\end{figure}
%
%

The shell correction does not only extract the ``shell effect'' from
the self--consistent binding energies, at the same time one obtains
the macroscopic part $\tilde{E}$ of the binding energy. 
A typical example is given in the left panel of Fig.~\ref{Fig:sc:mac}.
Owing to its optimization as a mass formula and its fit which includes 
also superheavy nuclei the YPE model can be expected to give a 
very good description of the macroscopic energy in this region and 
therefore can serve as reference. The curves for the various
interactions show considerable splitting which reflects of course 
the findings from Figs.~\ref{Fig:deltaE} and~\ref{Fig:deltaE2}.
This also confirms again the finding discussed above that the predictive 
power of an effective interaction for binding energy systematics is 
fairly independent of its predictive power for shell effects. 
On one hand SLy4 and NL3 give very similar values for $\tilde{E}$ 
as the YPE model (but remember the small errors in their trends
found in Fig.~\ref{Fig:deltaE2})
although they predict different magic numbers, on the other hand
those forces that give the best description of binding energies
independently on their disagreement on the next shell closure
(SkI3, SkI4, SkP, NL--Z2) give all very similar $\tilde{E}$ but
differ now significantly from the YPE model.

The right panel of Fig.~\ref{Fig:sc:mac} shows the macroscopic energy
$E_{\rm LDM}$ from the LDM expansion of the binding energy 
(\ref{eq:LDM}) using the 
values given in Table~\ref{Tab:NucMat}. Surprisingly $\tilde{E}$ and 
$E_{\rm LDM}$ differ on the order of 100 MeV. Only for $E_{\rm LDM}$
the ordering of the interactions is according to their value for 
\asym\ as one naively expects. This demonstrates that the leptodermous 
expansion with nuclear matter parameters does not work even for superheavy 
nuclei, finite--size effects are still important which also means that 
the use of nuclear matter parameters as pseudo--observables in the fit 
of \SC\ models might be dangerous.
%
%
\begin{figure}[t!]
\centerline{\epsfig{file=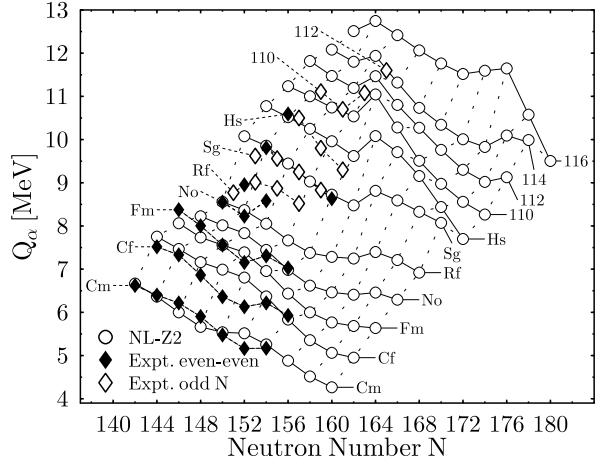}}
\caption{\label{Fig:Qa:even}
$Q_\alpha$ values of even-even nuclei calculated with NL-Z2 (open circles) 
compared with experimental values for even-even nuclei (filled diamonds) 
and odd--$N$ isotopes (open diamonds) of the heaviest even $Z$ elements. 
The data for odd--$N$ nuclei have to be handled carefully, some of these
might correspond to transitions involving excited states, and due to
blocking effects of the ground--state--to--ground--state values might
differ on the order of 500 keV from the systematics of $Q\alpha$ for
even--even nuclei.
Including data for nuclei with \mbox{$Z>116$} leads to overlapping 
curves, therefore those are omitted in the plot.
Data taken from\protect\cite{Ben00a}.
}
\end{figure}
%
%

Most of the recent new data on superheavy nuclei are $Q_\alpha$
values. Their systematics reflect all properties 
discussed above. As $\alpha$--decay chains have $I=$const.\ the
isoscalar channel of $\tilde{E}$ mainly determines the overall slope
of the $Q_\alpha$, while the isovector channel of $\tilde{E}$ 
mainly shifts the whole curves around. Shell effects bend
the curves locally, leading to kinks and peaks. A model
has to give a perfect description of all these properties
to reproduce experimental data throughout the superheavy region.
The quality of NL--Z2 for the $Q_\alpha$ is shown in Fig.~\ref{Fig:Qa:even}.
The overall description of the data is very good with the exception of 
some nuclei around \mbox{$Z=104$} where it overestimates a deformed shell
closure while the deformed \mbox{$N=152$} shell is shifted to \mbox{$N=150$}.
The latter is a problem from which virtually all \SC\ models 
suffer\cite{Cwi99a,Bue98a}.
%
%
\begin{figure}[t!]
\centerline{\epsfig{file=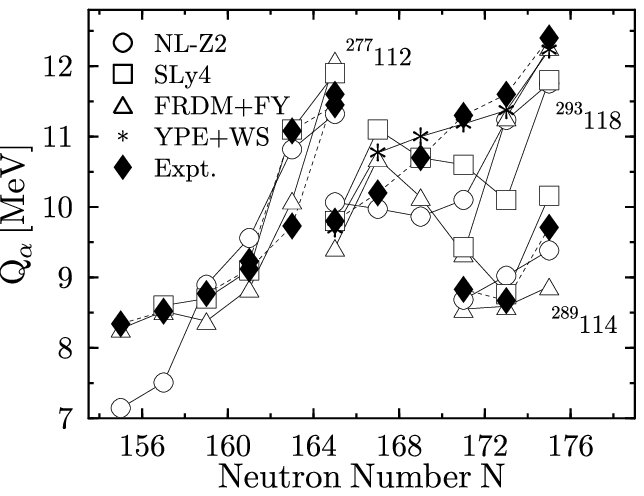}
            \epsfig{file=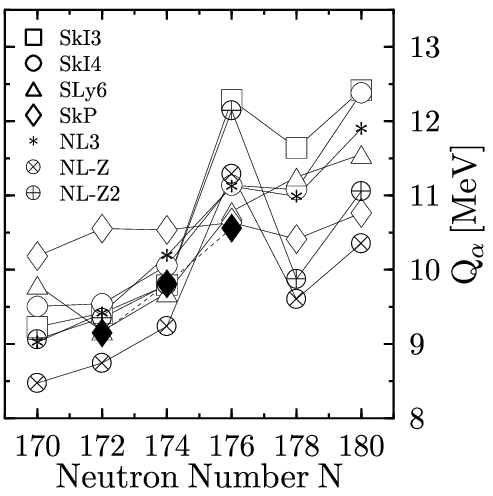}
}
\caption{\label{Fig:Qa:116}
Left panel:
Comparison of experimental and calculated $Q_\alpha$ values
for the decay chains of $^{277}_{165}112$, $^{289}_{175}114$,
and $^{293}_{175}118$, in the latter two cases following the
mass and charge assignment of the experimental groups.  
In the $^{277}_{165}112$ chain two distinct branches leading 
through different states of the intermediate nuclei are known. 
The calculated values from NL--Z2 and SLy4 connect the lowest 
states with positive parity in all cases (in the new chains 
only ${}^{289}_{175}114$ and ${}^{277}_{167}110$ are predicted 
by SLy4 and NL--Z2 to have ground states with negative parity), 
while the FRDM+FY and YPE+WS data are ground state to ground 
state values. 
Right panel: $Q_\alpha$ values of nuclei in the decay--chain 
${}^{300}_{180}120 \rightarrow {}^{296}_{178}118 \rightarrow
\ldots {}^{276}_{168}$Hs as predicted by the mean--field models
as indicated compared with preliminary experimental data. 
} 
\end{figure}
%
%

Fig.~\ref{Fig:Qa:116} compares some very recent data with calculations.
In view of the uncertainties, the \SC\ SLy4 and NL--Z2 give a very good
description of the data for the decay chain of $^{277}_{165}112$ (NL--Z2
agrees for \mbox{$Z > 104$} only as can be expected from 
Fig.~\ref{Fig:Qa:even}) and reproduces the \mbox{$N=162$} shell effect 
which cannot be seen in the FRDM+FY predictions. While all models
give similar predictions for this well--established chain, the spread
among the models is much larger for the new chains. All models with the 
exception of YPE+WS show spherical or deformed shells which cannot be seen
in the data.

Comparing predictions with the recent data for the even--even 
$^{292}_{176}$116 decay chain (which still have to be viewed as 
preliminary), see the right panel of Fig.~\ref{Fig:Qa:116}, it is 
most interesting that the data agree with calculated values from 
interactions which give different predictions for the spherical magic
numbers, i.e.\ SkI4 ($Z=114$, $N=184$), SLy6 ($Z=126$, $N=184$), and NL3 
($Z=120$, $N=172$). All other interactions show wrong overall trends of 
the $Q_\alpha$ or pronounced deformed shells in disagreement with the 
data or even both. The large difference between NL--Z and NL--Z2 is caused 
by their difference in symmetry energy although both forces predict the 
same shell structure. Once again all this demonstrates that predictions for 
spherical shell closures and binding energy systematics are fairly
independent. 

One of the reasons for the poor description of the new data is that 
-- independent on the actual location of the shell closures -- these nuclei 
are located in a region of transitional nuclei with very soft potential 
energy surfaces which amplify small differences in the shell structure. 
At the same time this adds a large uncertainty to the predictions as 
correlations beyond the mean--field level give a non--negligible 
contribution to binding energy differences which washes out shell 
effects visible in the results from mean--field calculations\cite{SHcoex}.
%
%
\section{Conclusions}
\label{sect:concl}
Combining the findings for average trends of binding energies and
single--particle spectra it has to be said that none of the current 
models is able to describe all available data.
Errors in the macroscopic part of the models have to be 
distinguished from errors in the shell structure. A good description 
of binding energy trends is not neccesarily an indicator for the 
predictive power concerning shell effects. Looking at
``macroscopic'' observables, a single parameterization of a model 
is not representative for the model while observables sensitive to 
shell effects reveal systematic differences between SHF and RMF.
The good overall description of spin--orbit splittings by the RMF
gives some preferrence for its predictions for magic numbers, but 
the undoubtedly missing isovector degrees of freedom in the 
RMF once included might also feed back to its predictions for superheavy
shell closures. More research in that direction has to be done.
%
%
\section*{Acknowledgments}
I would like to express my gratitute to all my collaborators with
whom the ideas and results presented here were develloped over the
years, in particular T.~B\"urvenich, 
S.~{\'C}wiok, P.--H.~Heenen, A.~Kruppa, J.~A.~Maruhn, W.~Nazarewicz, 
P.--G.~Reinhard, K.~Rutz, T.~Schilling, T.~Vertse, and W.~Greiner.
I am grateful for many fruitful discussions on experimental data
and their interpretation with S.~Hofmann, G.~M\"unzenberg, V.~Ninov, 
and Yu.~Ts.~Oganessian,
and I want to thank the organizers of the International Conference on 
``Fusion dynamics at the extremes'' for the kind invitation to present 
these results. A travel grant from the Herrmann Willkomm--Stiftung is 
gratefully acknowledged.
%
%

\end{document}